\documentclass[conference]{IEEEtran}

\usepackage{multirow}
\usepackage{graphicx}
\usepackage{subcaption}
\usepackage{hyperref}
\newtheorem{remark}{Remark}

\usepackage{amsmath}
\usepackage{amsfonts}           % blackboard math symbols  
\usepackage{amssymb}
\usepackage{nicefrac}           % compact symbols for 1/2, etc.
\usepackage{stmaryrd}

\usepackage{booktabs}
\usepackage{makecell}
\usepackage{comment}

\begin{document}

\title{BadSNN: Backdoor Attacks on Spiking Neural Networks via Adversarial Spiking Neuron}


\author{
    \IEEEauthorblockN{Abdullah Arafat Miah\IEEEauthorrefmark{1}, Kevin Vu\IEEEauthorrefmark{1}, Yu Bi\IEEEauthorrefmark{1}}\\
    \IEEEauthorblockA{\IEEEauthorrefmark{1}Department of Electrical, Computer, and Biomedical Engineering\\
    University of Rhode Island, Kingston, RI, USA\\
    \{abdullaharafat.miah, kevin\_vu029, yu\_bi\}@uri.edu}
}

\maketitle

\begin{abstract}

Spiking neural networks (SNNs) are energy-efficient counterparts of deep neural networks (DNNs) with high biological plausibility, as information is transmitted through temporal spiking patterns. The core element of an SNN is the spiking neuron, which converts input data into spikes following the leaky integrate-and-fire (LIF) neuron model. This model includes several important hyperparameters, such as the membrane potential threshold and membrane time constant. Both DNNs and SNNs have proven to be exploitable by backdoor attacks, where an adversary can poison the training dataset with malicious triggers and force the model to behave in an attacker-defined manner. Yet, how an adversary can exploit the unique characteristics of SNNs for backdoor attacks remains underexplored. In this paper, we propose \textit{BadSNN}, a novel backdoor attack on spiking neural networks that exploits hyperparameter variations of spiking neurons to inject backdoor behavior into the model. We further propose a trigger optimization process to achieve better attack performance while making trigger patterns less perceptible. \textit{BadSNN} demonstrates superior attack performance on various datasets and architectures while offering two key advantages over conventional data poisoning-based backdoor attacks: it opts out of conventional trigger poisoning and label flipping, and it demonstrates greater robustness to state-of-the-art backdoor mitigation techniques compared to existing attacks. Codes can be found at \url{https://github.com/SiSL-URI/BadSNN}. %The code will be made publicly available. 

% Our attack offers two key advantages over conventional data poisoning-based backdoor attacks: (1) it does not require conventional trigger poisoning and label flipping, and (2) consequently, it demonstrates greater robustness to state-of-the-art backdoor mitigation techniques compared to existing attacks.

\end{abstract}

\section{Introduction}
\label{sec:intro}

Deep neural networks (DNNs) have demonstrated significant performance across tasks ranging from computer vision \cite{deng2009imagenet, girshick2015fast}  to natural language understanding \cite{devlin2019bert, vaswani2017attention}. To  increase the performance of DNNs, models are becoming increasingly complex with large numbers of learnable weights and other parameters, consuming substantial energy for both training and inference \cite{dhar2020carbon}. To make models more energy-efficient and hardware-friendly, Spiking neural networks (SNNs) emerge as a promising alternative. SNNs operate with event-driven data similar to recurrent neural networks (RNNs), with information flowing through the model via discrete spikes rather than continuous values, making them energy-efficient and well-suited for edge devices \cite{maass1997thirdgen, ghoshdastidar2009snn, roy2019toward, fang2023spikingjelly}. As SNNs are increasingly deployed in safety-critical applications such as autonomous driving, medical devices, and surveillance systems, understanding their security vulnerabilities becomes a pressing concern.

SNNs can significantly reduce energy consumption and can be more robust to noise and different naturally occurring perturbations~\cite{kundu2021spike, lee2016training}. Additionally, SNNs can be implemented for real-time vision-based tasks like autonomous driving in place of DNNs as a more energy-efficient alternative~\cite{viale2021carsnn, wysoski2010evolving}. The main distinction between DNNs and SNNs is the presence of spiking neurons. The input to an SNN is processed over multiple timesteps, and at each timestep, these spiking neurons either fire a spike or remain silent, producing binary activations rather than continuous activations in DNNs~\cite{fang2023spikingjelly}. Due to this sparse nature, SNNs are optimized during backpropagation through spike timing-based plasticity~\cite{iyer2020classifying} or surrogate gradient-based optimization~\cite{wu2018spatio}. Another way of training SNNs is the ANN-SNN conversion approach, where the feature space of a pre-trained ANN is replicated into a corresponding
SNN~\cite{iyer2020classifying, han2020deep, sengupta2019going}.

Like DNNs, SNNs are also vulnerable to backdoor attacks \cite{gu2017badnets}. In a backdoor attack, the attacker is a malicious trainer or dataset provider who poisons the training dataset with triggers and forces the model to learn the association between a trigger and a target label. During inference, when the victim uses the backdoored model, the adversary can control the model's output by injecting the trigger into clean inputs. While backdoor attacks have been extensively studied in the DNN domain, with a wide range of attack strategies and corresponding defenses proposed in the literature \cite{gu2019badnets, chen2017targeted, nguyen2021wanet, gao2019strip, wu2021adversarial}, SNNs' vulnerabilities to backdoor attacks remain largely underexplored.

In terms of backdoor attacks on SNNs, most recent works primarily focus on neuromorphic data manipulation. Neuromorphic data is event-based data with a temporal dimension, where semantic information is represented through two different polarities (0/1). In Sneaky Spikes \cite{abad2023sneaky}, triggers are replicated from the BadNet \cite{gu2017badnets} attack and adapted for time-encoded and spike-encoded versions suitable for neuromorphic data. Because neuromorphic sensors such as DVS cameras naturally emit asynchronous spikes in response to changes in light intensity \cite{lichtsteiner2008dvs}, physical  triggers can take the form of flashes, strobes, or patterned bursts of light, which seamlessly blend with normal scene dynamics \cite{fang2024flashy}. However, this attack paradigm primarily focuses on neuromorphic data manipulation rather than exploiting vulnerabilities inherent to the SNN model itself. 

The unique characteristic of SNNs is that they contain spiking neurons~\cite{fang2021incorporating}, which convert input analog data into spike trains that flow through the network's layers. The behavior of these spiking neurons is governed by hyperparameters such as the membrane potential threshold ($V_{\text{thr}}$) and the membrane time constant ($\tau$), which directly control the number and timing of spikes generated for a given input. These hyperparameters are typically fixed by the model provider and are not subject to the same scrutiny during model inspection, creating an overlooked attack surface. To address this research gap and understand backdoor attacks from the perspective of SNNs' architectural uniqueness, we investigate the following research question in this paper: \textit{Can an attacker exploit the spiking neurons of SNNs to embed a backdoor attack?}

To answer this question, we propose \textit{BadSNN}, a novel backdoor
attack on spiking neural networks that exploits the
sensitivity of SNNs to the hyperparameters of spiking
neurons. Instead of traditional trigger poisoning, we employ
\textit{malicious spike poisoning} by tuning the
hyperparameters of spiking neurons during training. Such
approach fundamentally differs from existing SNN backdoor
attacks in two aspects: \textit{i)} it opts out of additional spikes injection and training data modification, and \textit{ii)} the
backdoor is embedded entirely through the model's intrinsic
spike generation mechanism rather than through extrinsic
trigger patterns. We also propose a trigger optimization
process to activate the backdoor during inference, generating
minimally perceptible perturbations that elevate spike
activity beyond the nominal range. Both static images and
neuromorphic data are evaluated to demonstrate the effectiveness and applicability of our proposed \textit{BadSNN}. Our contributions are summarized as follows:

\begin{itemize}
    \item We study the vulnerability of SNNs to the
    hyperparameters of spiking neurons and demonstrate that
    variations in the membrane potential threshold and
    membrane time constant can cause the model to treat
    in-distribution data as out-of-distribution, creating an
    exploitable attack surface.

    \item We propose \textit{BadSNN}, a novel backdoor attack on
    spiking neural networks through malicious spike poisoning
    by tuning the hyperparameters of spiking neurons during
    training, eliminating the need for any input data
    manipulation.

    \item We propose a trigger optimization process to
    generate minimally perceptible trigger perturbations that
    activate backdoor behavior during inference, leveraging a
    U-Net-based surrogate model trained with a combination of
    cosine similarity, adversarial, and weighted MSE losses.

    \item Through extensive experiments across four datasets
    (CIFAR-10, GTSRB, CIFAR-100, and N-MNIST) and three
    architectures (Spiking ResNet-19, Spiking VGG-16, and
    N-MNIST Net), we demonstrate the effectiveness of the
    proposed attack and its robustness against five
    state-of-the-art backdoor mitigation techniques,
    including pruning-based and fine-tuning-based defenses.
\end{itemize}
\section{Related works}
\label{sec:related-works}

Backdoor attacks have been well-studied across different deep
learning models, such as convolutional neural networks and
vision transformers~\cite{gu2017badnets, nguyen2021wanet,
cheng2024lotus, wang2022bppattack, gao2024dual,
miah2024noiseattack, subramanya2022backdoor, chan2022baddet},
language models~\cite{chen2021badnl, li2021hidden,
zhao2024exploring, sha2024prompt, miah2024exploiting}, and
graph neural networks~\cite{xi2021graph, zhang2024rethinking,
khan2026multi, zhang2024backdoor}. Early attacks such as
BadNets~\cite{gu2017badnets} demonstrated that injecting
poisoned samples with a static trigger can reliably cause
targeted misclassification.
Blend~\cite{liu2018trojaning} and
WaNet~\cite{saha2020hidden} extended this approach by
embedding subtler, more stealthy triggers. Recent works tend
to make triggers more stealthy by making them invisible in
both the spatial and frequency
domains~\cite{gao2024dual}, injecting triggers through
poisoned data
sub-partitioning~\cite{cheng2024lotus}, generating
imperceptible trigger perturbations through a surrogate
model~\cite{doan2021lira}, or using image quantization as
triggers~\cite{wang2022bppattack}. These works established
that poisoning during training poses a severe threat even
when the attacker controls only a small portion of the
dataset. In response, a wide range of defenses have been
proposed, such as pre-training defenses like anti-backdoor
learning~\cite{li2021anti}, trigger-inversion-based defenses
such as Neural Cleanse~\cite{wang2019neural},
poisoned-sample-guided detection methods like
STRIP~\cite{gao2019strip}, backdoor neuron pruning-based
mitigation techniques such as
ANP~\cite{wu2021adversarial} and
CLP~\cite{zheng2022data}, fine-tuning-based mitigation
techniques such as NAD~\cite{li2021neural} and
TSBD~\cite{lin2024unveiling}, and poisoned sample
purification-based techniques~\cite{yang2024sampdetox,
miah2026lite, shi2023black}.

Backdoor attacks on SNNs are relatively recent but are rapidly gaining 
attention. One important study, the Sneaky Spikes framework 
\cite{abad2024sneakyspikes}, showed that injecting a small fraction of poisoned 
temporal event bursts into training data can embed highly effective backdoors 
in surrogate-trained SNNs. Beyond digital poisoning, physical neuromorphic 
backdoors exploit the properties of event-based sensors. Flashy Backdoor 
\cite{fang2024flashy} demonstrated that real-world DVS recordings can be 
compromised with timed light flashes or strobing patterns that blend naturally 
into the event stream. Such physical triggers remain effective under varying 
lighting and motion conditions, underscoring the real-world risks associated 
with neuromorphic sensing. Other works show that the sparseness of spiking 
activity and conditional firing dynamics make backdoors resistant to 
traditional defense strategies. Data-poisoning attacks targeting supervised SNN 
learning have been shown to persist even after fine-tuning or pruning 
\cite{jin2024snnbackdoor}. However, existing backdoor attacks cannot be easily 
adapted to static images because they primarily focus on exploiting the 
temporal dimension of neuromorphic data.

\section{Methodology} \label{sec:methodology}
\subsection{Threat Model}

We adopt the threat model conventionally used in state-of-the-art backdoor attacks \cite{gu2019badnets,nguyen2021wanet,cheng2024lotus}. The adversary is assumed to have white-box access to the victim model and full control over the training process. Their objective is to maximize the attack success rate while preserving the model’s clean utility. In contrast to conventional approaches that perform data poisoning using mislabeled samples, the adversary in our setting manipulates the hyperparameters of spiking neurons during training to embed the backdoor. Furthermore, unlike traditional methods that predefine the trigger function prior to backdoor training, the attacker constructs and optimizes the trigger function after the training phase.

\subsection{Preliminaries}

SNNs are the counterpart of conventional DNNs, where the core unit is the spiking neuron, which converts input data into spike trains to mimic biologically plausible neurons. The most popular model for simulating biological neurons in SNNs is the Leaky Integrate-and-Fire (LIF) model~\cite{maass1997networks}. LIF neurons take inputs from one layer of an SNN model and fire a spike to the next layer when the membrane potential of the neurons exceeds a threshold. The working principle of an LIF neuron can be described by Equation~\ref{eq:lif_eq}.

\begin{equation} \label{eq:lif_eq}
\begin{aligned}
\tau \frac{dV(t)}{dt} &= -[V(t) - V_{rest}] + RI(t)\\
S(t) &= \begin{cases}
1, & \text{if } V(t) \geq V_{thr} \\
0, & \text{if } V(t) < V_{thr}
\end{cases}
\end{aligned}
\end{equation}

\noindent where $V(t)$ is the membrane potential, $I(t)$ is the input to the neuron at time $t$, $R$ is the membrane resistance, $\tau$ is the membrane time constant, and $V_{rest}$ is the resting potential. 

When the membrane potential $V(t)$ exceeds a certain threshold $V_{thr}$, the neuron fires a spike $S(t)$ at time $t$ followed by a reset value $V_{reset}$ for $V(t)$. 
% The input to the spiking neuron in a particular layer $l$ can be described as $I^l(t) = \mathbf{W}^lS^{l-1}(t) + b^l$, where $\mathbf{W}$ is the weight matrix of the convolutional layers and $b$ is the bias term. Thus, a typical block of an SNN consists of convolutional layers, pooling layers, normalization layers, and a LIF neuron layer. 
Therefore, two main hyperparameters can dictate the LIF neurons: the membrane potential threshold $V_{thr}$ and the membrane time constant $\tau$. Although both hyperparameters in most cases are chosen by the model provider, there are approaches that make them learnable, such as in parametric LIF~\cite{fang2021incorporating}, where the membrane time constant $\tau$ is learned alongside the weights for improved temporal representation. All of this information propagation is performed simultaneously across multiple timesteps and averaged at the end to construct the final output layer. Due to the spiking neurons, SNNs can operate on sparse spike events, which consume significantly less power than traditional neural networks. However, their performance heavily depends on these hyperparameters, which creates opportunities for malicious manipulation.

\begin{comment}

%horizontal spacing

\begin{figure}[t]
    \centering
    \begin{subfigure}[b]{\linewidth}
        \centering
        \includegraphics[width=0.6\linewidth]{Figure/accuracy_vs_threshold_cifar10_resnet19.pdf}
        \caption{Test accuracy}
        \label{fig:accuracy_vs_threshold}
    \end{subfigure}

    \begin{subfigure}[b]{\linewidth}
        \centering
        \includegraphics[width=0.6\linewidth]{Figure/spikes_vs_threshold_cifar10_resnet19_all_layers_20251113_221614.pdf}
        \caption{Average spikes per sample}
        \label{fig:spike_vs_threshold}
    \end{subfigure}
    
    \caption{Effect of membrane potential threshold ($V_{thr}$) on ResNet-19 performance for CIFAR-10: (a) test accuracy degradation and (b) average spike count per sample across all LIF neuron layers.}
    \label{fig:threshold_effects}
\end{figure}
    
\end{comment}

\begin{figure}[t]
    \centering
    \begin{subfigure}[b]{0.48\linewidth}
        \centering
        \includegraphics[width=\linewidth]{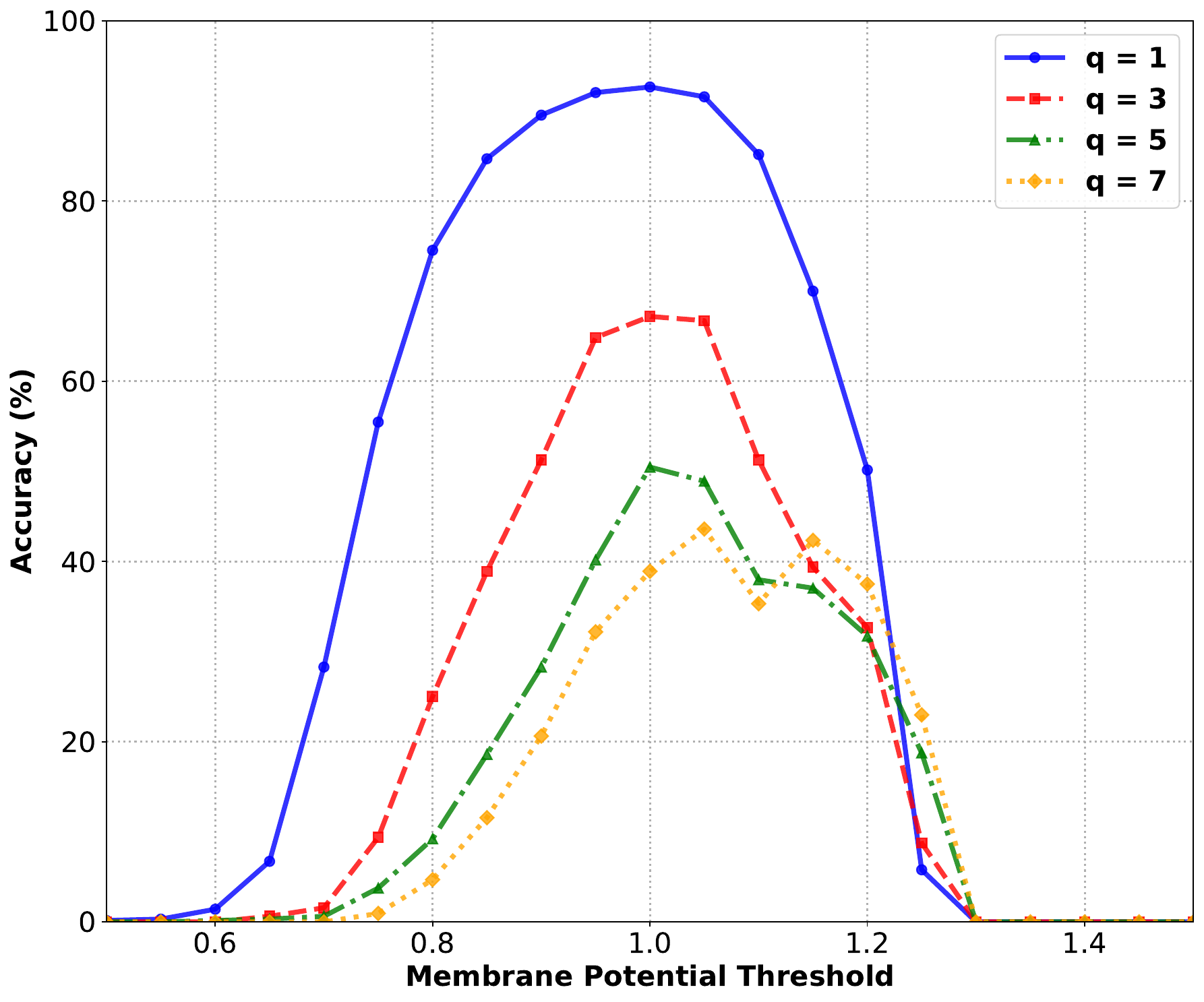}
        \caption{Test accuracy}
        \label{fig:accuracy_vs_threshold}
    \end{subfigure}
    \hfill
    \begin{subfigure}[b]{0.48\linewidth}
        \centering
        \includegraphics[width=\linewidth]{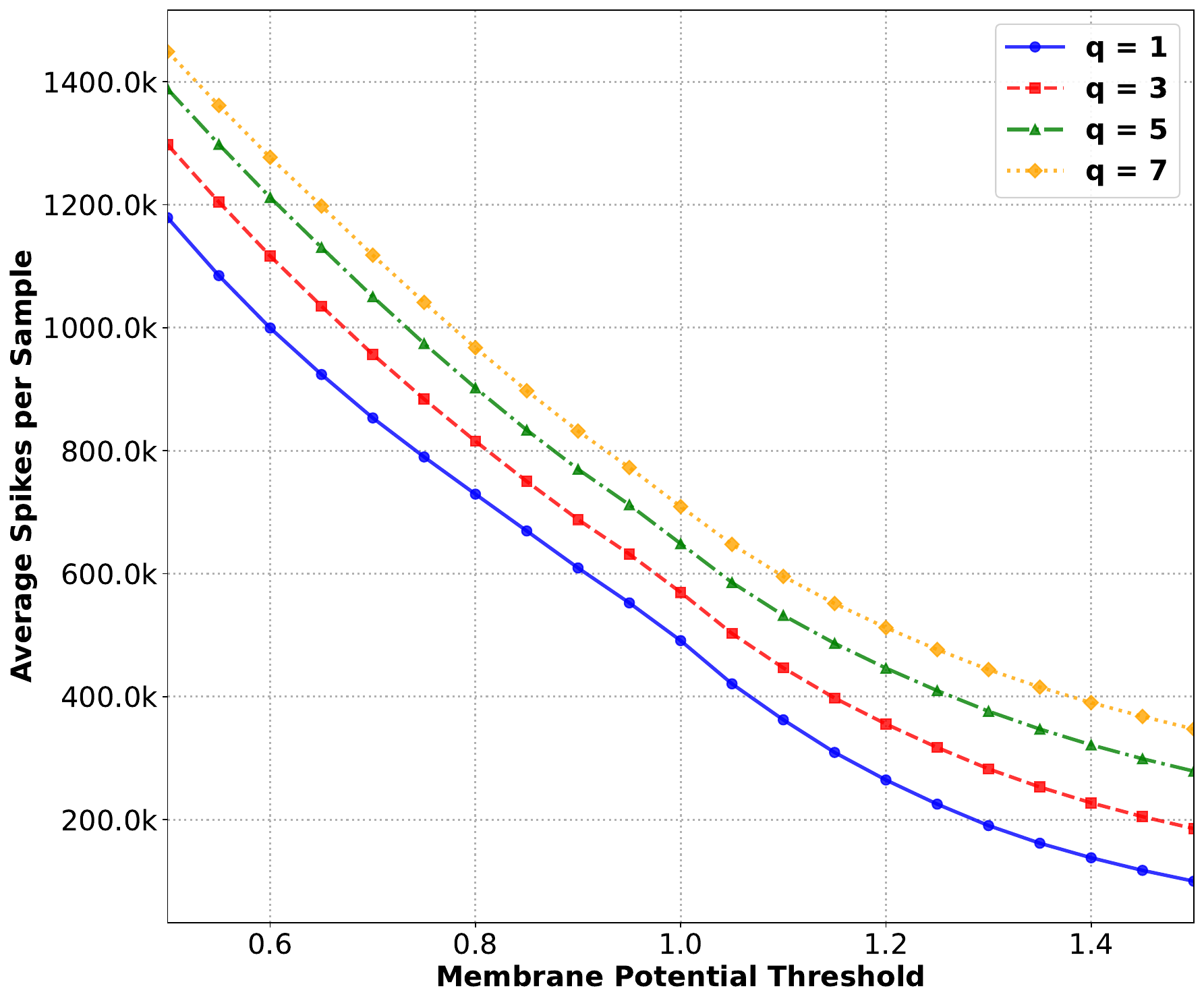}
        \caption{Average spikes per sample}
        \label{fig:spike_vs_threshold}
    \end{subfigure}
    
    \caption{Effect of membrane potential threshold ($V_{thr}$) on spiking ResNet-19 performance for CIFAR-10: (a) test accuracy degradation and (b) average spike count per sample across all LIF neuron layers.}
    \label{fig:threshold_effects}
\end{figure}

Specifically, both $V_{thr}$ and $\tau$ control the number of spikes $\mathcal{N}_{spike}$ generated in the model for any given input sample. Given that SNN accuracy is highly tied with $\mathcal{N}_{spike}$, varying $V_{thr}$ might intuitively fluctuate model accuracy and performance. To demonstrate this, we use a spiking ResNet-19 \cite{li2023seenn} architecture trained on the CIFAR-10 dataset with $V_{thr} = 1$ and $\tau = 0.5$. We then evaluate the testset accuracies for various values of $V_{thr}$, as shown in Figure~\ref{fig:accuracy_vs_threshold}, where the accuracy degrades significantly when $V_{thr}$ deviates from its nominal value ($V_{thr} = 1$, $\tau = 0.5$). As spike $S(t)$ is proportional to the input $I(t)$ provided in Equation~\ref{eq:lif_eq}, the input values including pixel intensities and their convolved/pooled representations can heavily influence the SNN's behavior. 
To further test the above observation, we then apply an element-wise nonlinear power transformation function $f_{q}: x_i \mapsto x_i^{q}$ to the input sample, where $q$ is power ratio. Such transformation ($q > 1$) consistently generates a larger number of spikes compared to the original image ($q = 1$), causing the SNN to treat the transformed input as out-of-distribution data, as evidenced in both Figure~\ref{fig:accuracy_vs_threshold} and Figure~\ref{fig:spike_vs_threshold}. 

\begin{remark} \label{remark: remark_1}
This phenomenon suggests that there exists a feature space for a given SNN where input samples generate elevated spike counts and are subsequently considered as out-of-distribution. If the spiking neuron hyperparameters can be tuned in a way that the model exhibits a bias toward a target label for these out-of-distribution samples, nonlinear transformations on input samples can be leveraged to launch a backdoor attack.
\end{remark}

\subsection{Proposed Attack}

\begin{figure}
    \centering
    \includegraphics[width=\linewidth]{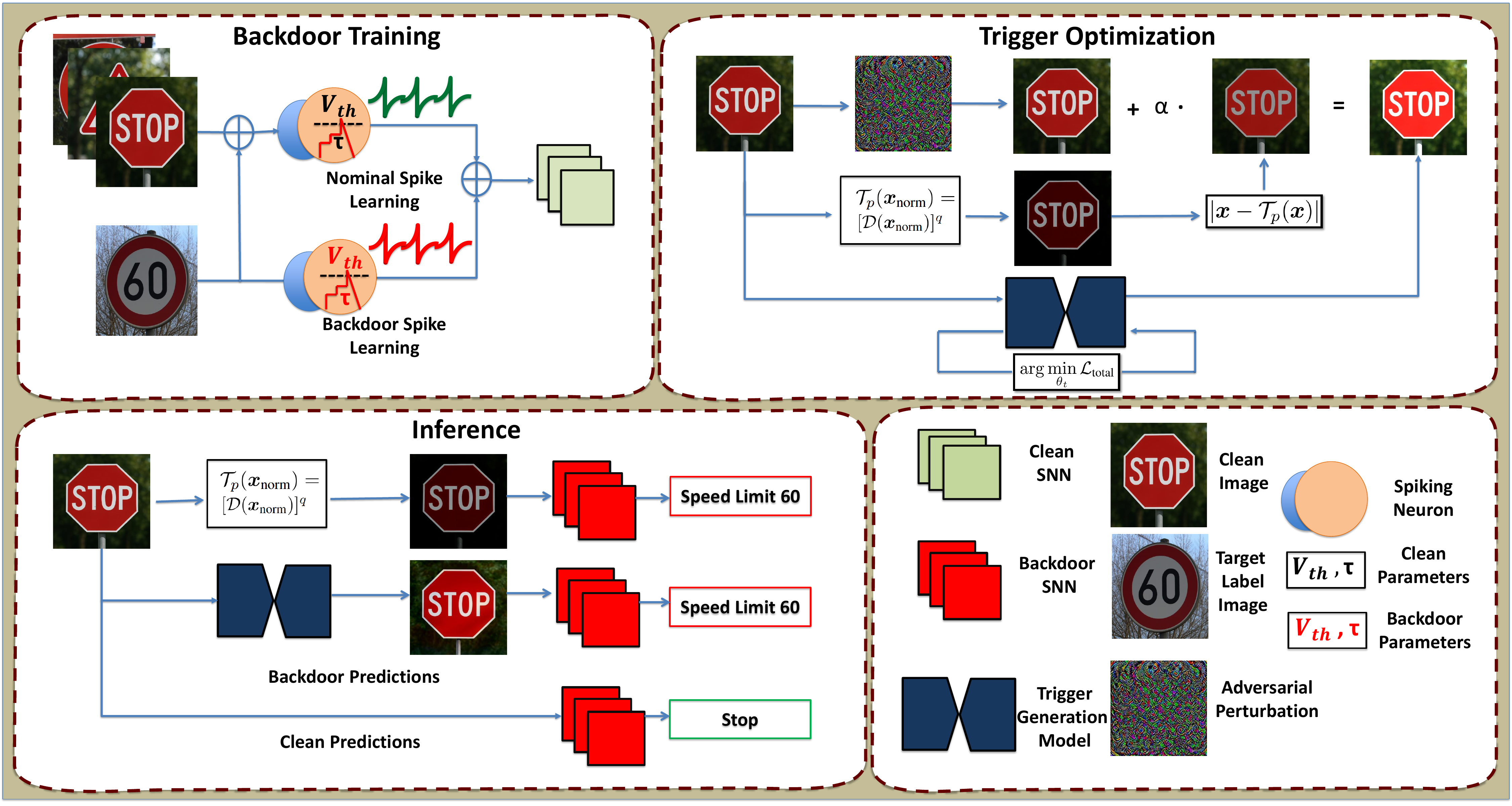}
    \caption{Overview of the proposed \textit{BadSNN}. \textit{BadSNN} has three major steps: Backdoor Training, Trigger Optimization, and Inference. During Backdoor Training, the adversary manipulates the hyperparameters of the spiking neurons in the target clean SNN to perform dual spike learning. In the Trigger Optimization step, the weights of a trigger generation model are optimized to generate minimally perceptible trigger perturbations that induce out-of-distribution spike counts. In the Inference step, the victim uses the backdoored SNN, which produces clean outputs for clean samples but produces adversarially defined outputs for triggered samples.}
    \label{fig:overview}
\end{figure}

Inspired by \textit{Remark 1}, we propose a novel backdoor attack scheme on spiking neural networks, namely \textit{BadSNN}, where we deviate from the conventional poisoning-based backdoor attacks. Instead, we follow a variable LIF hyperparameter tuning approach to embed backdoor behavior in the model. 
The affected SNN treats samples that generate spikes outside the normal spike range as in-distribution data belonging to the target label. 
We refer to this method as malicious spike poisoning. After backdoor training, we propose a trigger optimization approach, where we employ a trigger surrogate model that generates trigger perturbations to fool the model into predicting its input as the target label. The attack overview of the proposed
\textit{BadSNN} is given in Figure \ref{fig:overview}.

\subsubsection{Backdoor Training}

Consider a spiking neural network ($F$) consisting of spiking neurons $\mathcal{S}(V_{\text{thr}}, \tau)$ following the LIF model as described in Equation \ref{eq:lif_eq}, with a fixed number of time steps $T$. Let $D = \{(\boldsymbol{x}_i, y_i)\}_{i=1}^N$ denote the training set containing $N$ samples, where $\boldsymbol{x}_i \in \mathcal{X}$ represents the input sample space and $y_i \in \mathcal{Y}$ represents the output classification space. The network $F$ is trained on $D$ to learn the mapping $F_{\theta} : \mathcal{X} \to \mathcal{Y}$, where $\theta$ denotes the learnable parameters (weights) of $F$. 

To facilitate backdoor training, we partition the training set into three subsets. Let $y_t$ denote the target label, and let $D_n$ and $D_t$ represent the subsets of $D$ corresponding to non-target and target label classes, respectively. We further partition $D_t$ into $D_t^c$ and $D_t^p$, where the poisoning ratio is defined as $P = \frac{|D_t^p|}{|D|}$. We define nominal hyperparameters of $\mathcal{S}$ as $V_{\text{thr}}^n$ and $\tau^n$, and malicious hyperparameters as $V_{\text{thr}}^t$ and $\tau^t$. During backdoor training, we employ two spiking neuron configurations: $\mathcal{S}_n = \mathcal{S}(V_{\text{thr}}^n, \tau^n)$ and $\mathcal{S}_t = \mathcal{S}(V_{\text{thr}}^t, \tau^t)$ to represent the nominal and malicious spike generation configurations, respectively. The backdoor training objective of minimizing the backdoor loss $\mathcal{L_B}$ is formalized in Equation \ref{eq:backdoor_learning}, where $\theta_c$ and $\theta_b$ denote the clean and backdoor weights, respectively.

\begin{equation} \label{eq:backdoor_learning}
\resizebox{\linewidth}{!}{$\displaystyle
\arg\min_{\theta = \theta_c \cup \theta_b} \Big[ 
\underbrace{\sum_{D_n \cup D_t^c} \mathcal{L}(F(x_i; \theta_c; \mathcal{S}_n), y_i)}_{\text{Nominal Spikes Learning}}
+ \underbrace{\sum_{D_t^p} \left[\mathcal{L}(F(x_i; \theta_c; \mathcal{S}_n), y_i) + \mathcal{L}(F(x_i; \theta_b; \mathcal{S}_t), y_i)\right]}_{\text{Dual Spikes Learning}} \Big]
$}
\end{equation}

The dual spike learning methodology enforces a subset of samples from the target label class to be trained with both $\mathcal{S}_n$ and $\mathcal{S}_t$, enabling the model to learn the distinction between nominal and malicious spikes. Notably, we only tune the spiking neuron's hyperparameters without adding any trigger function to generate poisoned samples during backdoor training. 
% Furthermore, we do not perform label flipping during training, thereby adhering to the clean-label attack framework. 
When selecting malicious hyperparameters $V_{\text{thr}}^t$ and $\tau^t$, the adversary must satisfy two criteria: \textbf{i)} With malicious hyperparameters, the SNN expects to regard in-distribution data as out-of-distribution (i.e., exhibiting low accuracy), and \textbf{ii)} Malicious hyperparameters need to be configured distant from the nominal hyperparameters preventing their interference with the SNN's clean model performance. 
% \textbf{iii)} It should be possible to identify attack hyperparameters where both clean performance and backdoor performance are preserved.

\subsubsection{Trigger Optimization}

Based on the previous analysis, we observe that there exist nonlinear transformations that elevate spike activity and deceive the model into associating these transformed images with the target label. This occurs because the model has been trained biasing toward the target label for certain spike patterns that deviate from the nominal range. We initialize our trigger function ($\mathcal{T}_p$) with the nonlinear power transformation defined in Equation \ref{eq:power_transformation}, where $\mathcal{D}(\boldsymbol{x}_{\text{norm}}) = \boldsymbol{x}_{\text{norm}} \odot \boldsymbol{\sigma} + \boldsymbol{\mu}$, with $\boldsymbol{\mu}$ and $\boldsymbol{\sigma}$ representing the mean and standard deviation of the normalization process, respectively.

\begin{equation} \label{eq:power_transformation}
\mathcal{T}_p(\boldsymbol{x}_{\text{norm}}) = [\mathcal{D}(\boldsymbol{x}_{\text{norm}})]^q
\end{equation}

However, $\mathcal{T}_p$ has two major limitations: (1) the perturbations generated by this transformation may compromise the semantic information of the input samples, and (2) the perturbations may be perceptible to human observers due to their noticeable differences from the original images. To address these limitations, we propose a trigger optimization process to generate minimally perceptible perturbations that can fool the classifier toward the target label.

Let $\mathcal{T}_o$ denote a conditional image transformation function with learnable parameters $\theta_t$. After training, $\mathcal{T}_o$ should generate trigger perturbations that are less perceptible than those produced by $\mathcal{T}_p$ while offering stronger attack performance. To solve this complex optimization problem, we leverage the trained backdoor SNN to search for minimal perturbations for each training sample, which $\mathcal{T}_o$ will subsequently learn to generate.

Given a training sample $\boldsymbol{x}$, we use the backdoor SNN to generate adversarial examples by adding imperceptible noise that pushes the sample beyond its original decision boundary. Less intensive nonlinear transformations are expected to shift predictions toward the target label for adversarial examples, since samples with adversarial perturbation already lie outside their original decision boundary. Let $\boldsymbol{\delta}_{\text{DF}}$ denote an adversarial example generator following the DeepFool algorithm \cite{moosavi2016deepfool}. We perform an adaptive blending operation to obtain a minimally transformed image $\boldsymbol{x}^{\text{blend}}$ that will be classified as the target label by the backdoor SNN. The adaptive blending strategy is formalized in Equation \ref{eq:adaptive_blend}, where $\alpha$ represents an adaptive blending ratio determined through grid search over $k$ candidates.

\begin{equation} \label{eq:adaptive_blend}
\begin{aligned}
\boldsymbol{x}^{\text{blend}} &= (1 - \alpha) \cdot |\boldsymbol{x} - \mathcal{T}_p(\boldsymbol{x}) | + \alpha \cdot (\boldsymbol{x} + \boldsymbol{\delta}_{\text{DF}}(\boldsymbol{x})) \\
\Delta_{\boldsymbol{x}} &= |\boldsymbol{x} - \boldsymbol{x}^{\text{blend}}|
\end{aligned}
\end{equation}

For a given $\boldsymbol{x}$, $\mathcal{T}_o$ is trained to generate $\Delta_{\boldsymbol{x}}$. The total training loss $\mathcal{L}_{\text{total}}$ of $\mathcal{T}_o$ is a weighted combination of three learning objectives given in Equation \ref{eq:trigger_loss}:

\begin{equation} \label{eq:trigger_loss}
\arg\min_{\theta_t} \mathcal{L}_{\text{total}} = \lambda_1 \mathcal{L}_{\text{sim}} + \lambda_2 \mathcal{L}_{\text{adv}} + \lambda_3 \mathcal{L}_{\text{wmse}}
\end{equation}

Here, $\mathcal{L}_{\text{sim}}$ corresponds to the cosine similarity loss, which enforces structural alignment between the generated perturbation and the target perturbation; $\mathcal{L}_{\text{adv}}$ is the adversarial loss function that ensures the perturbed image is successfully misclassified to the target label; and $\mathcal{L}_{\text{wmse}}$ is the weighted mean squared error (MSE) loss that matches the magnitude and spatial distribution of the generated perturbation with the target perturbation. The weight coefficients $\lambda_1, \lambda_2, \lambda_3$ are set to 1.0, 0.1, and 1.0, respectively. The formulations of these loss components are described in Equation \ref{eq:trigger_indv_loss}, where $\langle \cdot, \cdot \rangle$ denotes the inner product, $\text{CE}(\cdot, \cdot)$ is the cross-entropy loss, and $C$, $H$, $W$ are the channel, height, and width dimensions of the input samples, respectively. We utilize the U-Net architecture \cite{ronneberger2015u} as our trigger model $\mathcal{T}_o$.

\begin{equation} \label{eq:trigger_indv_loss}
\resizebox{\linewidth}{!}{$
\begin{aligned}
\mathcal{L}_{\text{sim}} &= 1 - \frac{\langle \mathcal{T}_o(\boldsymbol{x}, y_t; \theta_t), \Delta_{\boldsymbol{x}} \rangle}{\|\mathcal{T}_o(\boldsymbol{x}, y_t; \theta_t)\|_2 \|\Delta_{\boldsymbol{x}}\|_2} \\
\mathcal{L}_{\text{adv}} &= \text{CE}\left(F_{\theta}(\boldsymbol{x} + \mathcal{T}_o(\boldsymbol{x}, y_t; \theta_t); \mathcal{S}_n), y_t\right)\\
\mathcal{L}_{\text{wmse}} &= \frac{1}{C \cdot H \cdot W} \sum_{c,h,w} w_{c,h,w} \cdot \left(\mathcal{T}_o(\boldsymbol{x}, y_t; \theta_t)_{c,h,w} - \Delta_{\boldsymbol{x},c,h,w}\right)^2
\end{aligned}
$}
\end{equation}

\subsubsection{Extension to Neuromorphic Data} 

Neuromorphic data is event-based data recorded using neuromorphic hardware, consisting of discrete spikes rather than the continuous analog values of static images. Therefore, our proposed trigger optimization approach cannot be directly applied to neuromorphic data. However, given \textit{Remark 1}, \textit{BadSNN} can remain true for neuromorphic data in the case that if we apply a transformation function that elevates the number of spikes, the model will classify the input to the target label. We propose a trigger function $\mathcal{T}_s$ that introduces temporally-varying random perturbations to the spike frames, as illustrated in Equation \ref{eq:neuromorphic_trigger}. Here, $\boldsymbol{\epsilon}$ is a random perturbation tensor with elements drawn from a uniform distribution $\mathcal{U}(-\beta, \beta)$, where $\beta$ controls the trigger intensity. In our experiments, we set $\beta = 0.03$. The $\text{clip}(\cdot, 0, 1)$ operation ensures the perturbed data remains within valid spike frame boundaries. As this transformation introduces random spikes into the input, the total spike activity generated by the transformed data exceeds the nominal range, thereby activating the backdoor.

\begin{equation} \label{eq:neuromorphic_trigger}
\mathcal{T}_s(\boldsymbol{x}; \beta) = \text{clip}(\boldsymbol{x} + \boldsymbol{\epsilon}, 0, 1)
\end{equation}

\subsubsection{Inference} 

During inference with the backdoored SNN, we can activate the backdoor by applying either $\mathcal{T}_p$ or $\mathcal{T}_o$ for static images, and $\mathcal{T}_s$ for neuromorphic data. However, a critical question arises: what should be the hyperparameters of the spiking neurons ($\mathcal{S}$)? If we use nominal hyperparameters ($V_{\text{thr}}^n$ and $\tau^n$), the clean task performance will be higher but attack effectiveness will be lower. Conversely, if we use malicious hyperparameters ($V_{\text{thr}}^t$ and $\tau^t$), the attack effectiveness will be highest but the clean task performance will be compromised. Therefore, the adversary must choose attack hyperparameters ($V_{\text{thr}}^a$ and $\tau^a$) that lie between the nominal and malicious hyperparameters and satisfy the conditions in Equation \ref{eq:inference}, where $y_g$ denotes the ground truth label of input sample $\boldsymbol{x}$ and $\mathcal{T} \in \{\mathcal{T}_p, \mathcal{T}_o, \mathcal{T}_s\}$.

\begin{equation} \label{eq:inference}
\begin{aligned}
    F_{\theta}(\boldsymbol{x}, \mathcal{S}(V_{\text{thr}}^a, \tau^a)) &= y_g \\
    F_{\theta}(\boldsymbol{x} + \mathcal{T}(\boldsymbol{x}), \mathcal{S}(V_{\text{thr}}^a, \tau^a)) &= y_t
\end{aligned}
\end{equation}

\section{Experiments} \label{sec:experiment}

\subsection{Experimental Settings}

\subsubsection{Datasets, Models, and Training Details} \label{subsubsec:datasets_models}

To investigate the effectiveness of the proposed \textit{BadSNN}, we design our experiments by incorporating four popular datasets consisting of static images and neuromorphic data: CIFAR-10 \cite{krizhevsky2009learning}, GTSRB \cite{stallkamp2011german}, CIFAR-100 \cite{krizhevsky2009learning}, and N-MNIST \cite{orchard2015converting}. CIFAR-10 is a widely used dataset which consists of 60,000 color images with 32 $\times$ 32 resolution across 10 object categories. CIFAR-100 is an extended version of CIFAR-10, where 60,000 images are distributed across 100 object categories with the same resolution as CIFAR-10. GTSRB is a real-world traffic sign classification dataset containing over 50,000 images distributed across 43 traffic sign categories. For CIFAR-10, CIFAR-100, and GTSRB, we split the datasets in an 80\%/20\% training and testing manner. N-MNIST is a neuromorphic version of the classic MNIST dataset \cite{orchard2015nmnist}, which was generated by recording MNIST digit stimuli displayed on an LCD screen using a Dynamic Vision Sensor (DVS) camera. It has 10 class categories with 60,000 / 10,000 train-test split. All the reported evaluation metrics are measured on the testing set.

We use Spiking ResNet-19 \cite{li2023seenn} to classify the CIFAR-10 dataset, Spiking VGG-16 \cite{li2023seenn} to classify the GTSRB and CIFAR-100 datasets, and N-MNIST Net \cite{abad2023sneaky} to classify the neuromorphic N-MNIST dataset. The Spiking ResNet-19 adopts a [3, 3, 2] block configuration, where each residual block contains two 3$\times$3 convolutional layers, each followed by temporal batch normalization (tdBN) and a LIF spiking neuron, along with residual skip connections. The Spiking VGG-16 employs a straightforward sequential stack consisting of five blocks, each containing 3$\times$3 convolutional layers followed by tdBN and a LIF spiking neuron, with average pooling for spatial downsampling. The N-MNIST Net is a two-layer convolutional model with LIF spiking neurons after each layer. For the trigger generation model, we employ a U-Net architecture \cite{ronneberger2015u}. The encoder consists of three convolutional blocks, each containing two 3$\times$3 convolutional layers followed by 2D batch normalization and ReLU activation, with max pooling for downsampling between blocks. The decoder mirrors the encoder structure but employs bilinear interpolation for upsampling and concatenates skip connections from the corresponding encoder stages. The output perturbation is bounded within $[-\epsilon, \epsilon]$ via Tanh scaling. We set $\epsilon = 0.3$ in all experiments.

For training the SNNs, we use the direct training methodology with fixed timesteps $ T = 4 $ for the Spiking ResNet-19 and VGG-16. All spiking neurons in our experiments employ the LIF neuron model. During backpropagation, we replace the non-differentiable Heaviside step function with the DSPIKE \cite{li2021differentiable} surrogate gradient method. We employ Stochastic Gradient Descent (SGD) as the optimizer.

\subsubsection{Attack and Defense Baselines}

We compare our proposed \textit{BadSNN} with three state-of-the-art conventional backdoor attacks: BadNet \cite{gu2017badnets}, Blend \cite{chen2017targeted}, and WaNet \cite{nguyen2021wanet}. BadNet uses a pattern or patch as a trigger to activate the backdoor. In our experiment, we use a 6$\times$6 checkerboard pattern as the trigger. For the Blend attack, a trigger image is blended with a clean image to activate the backdoor. In our experiment, we employ a blending ratio of 0.1 for the trigger. WaNet uses grid-based warping to generate stealthy triggers. In our experiment, we utilize a warping strength of 0.5 for WaNet. During backdoor training, a 5\% poisoning ratio is applied to all baseline attacks.

Additionally, we evaluate five state-of-the-art backdoor defense methods: Fine-Tuning, CLP \cite{zheng2022data}, ANP \cite{wu2021adversarial}, TSBD \cite{lin2024unveiling}, and NAD \cite{li2021neural}. In vanilla Fine-Tuning defense, the model is trained with a clean dataset to reduce the attack effect. CLP and ANP are pruning-based backdoor mitigation techniques. Both techniques attempt to identify backdoor-related neurons that are dominant for backdoor-related tasks and prune them to reduce the backdoor effect while preserving clean utility. In our experiment, for CLP we set a fixed threshold of $u=3$, and for ANP we set the perturbation budget to 0.4 and the hyperparameter $\alpha = 0.5$, and assume access to 5\% clean test data. TSBD and NAD are two advanced fine-tuning-based backdoor mitigation techniques. TSBD performs activeness-aware fine-tuning instead of vanilla fine-tuning. NAD uses a guided fine-tuning approach through neural attention distillation.

\vspace{-5mm}

\subsubsection{Evaluation Metrics}

Before introducing the evaluation metrics, for clarity, we reiterate some of the important notations that will be varied to calculate different evaluation metrics. During inference, the hyperparameters of the spiking neuron $\mathcal{S}$ can be set to either nominal configurations ($V_{\text{thr}}^n$ and $\tau^n$) or attack configurations ($V_{\text{thr}}^a$ and $\tau^a$). For static images, we have two trigger functions that can generate trigger perturbations, namely $\mathcal{T}_p$ and $\mathcal{T}_o$. For neuromorphic data, we have a single trigger function $\mathcal{T}_s$. To demonstrate the effectiveness and robustness of the proposed \texttt{BadSNN}, we employ four evaluation metrics: (1) Clean CA: the accuracy of the clean task for clean models without any backdoor training, (2) Base CA: the accuracy of the clean task for the backdoor model under nominal hyperparameter configurations ($V_{\text{thr}}^n$ and $\tau^n$) of the spiking neurons ($\mathcal{S}$), (3) CA: the accuracy of the clean task for the backdoor model under attack hyperparameter configurations ($V_{\text{thr}}^a$ and $\tau^a$) of the spiking neurons ($\mathcal{S}$), and (4) Attack Success Rate (ASR): the proportion of samples classified as the target label when triggered. Two types of ASR are evaluated for \textit{BadSNN}: $ASR_p$ illustrates the ASR when triggered with $\mathcal{T}_p$ or $\mathcal{T}_s$, while $ASR_o$ describes the ASR when triggered with $\mathcal{T}_o$. For clarity, we provide the mathematical definition of each evaluation metric in Equations~\ref{eq:base_ca} to \ref{eq:asro}, where $N$ is the total number of testing samples, $y_g$ denotes the ground truth, $y_t$ denotes the target label, and $\delta(a,b) = 1$ if $a = b$ and $\delta(a,b) = 0$ otherwise.
\begin{align}
    \text{Base CA} &= \frac{\sum_{i=1}^{N} \delta\bigl( F_{\theta}(\boldsymbol{x}, \mathcal{S}(V_{\text{thr}}^n, \tau^n)), y_g\bigr)}{N} \label{eq:base_ca}\\
    \text{CA} &= \frac{\sum_{i=1}^{N} \delta\bigl( F_{\theta}(\boldsymbol{x}, \mathcal{S}(V_{\text{thr}}^a, \tau^a)), y_g\bigr)}{N} \label{eq:ca}\\
    ASR_p &= \frac{\sum_{i=1}^{N} \delta\bigl( F_{\theta}\bigl(\mathcal{T}_p(\boldsymbol{x}) \lor \mathcal{T}_s(\boldsymbol{x}), \mathcal{S}(V_{\text{thr}}^a, \tau^a)\bigr), y_t\bigr)}{N} \label{eq:asrp}\\
    ASR_o &= \frac{\sum_{i=1}^{N} \delta\bigl( F_{\theta}\bigl(\mathcal{T}_o(\boldsymbol{x}), \mathcal{S}(V_{\text{thr}}^a, \tau^a)\bigr), y_t\bigr)}{N} \label{eq:asro}
\end{align}

\subsection{Attack Effectiveness Analysis}

%To demonstrate the effectiveness of \textit{BadSNN}, we summarize the results for different datasets and models when they are trained with given malicious hyperparameters ($V_{\text{thr}}^t$ and $\tau^t$) and evaluated under different attack hyperparameters ($V_{\text{thr}}^a$ and $\tau^a$) in Table \ref{tab:main_results}. An acceptable accuracy degradation of Base CA from Clean CA can be observed across all datasets. The CA and ASR vary greatly for different $V_{\text{thr}}^a$ and $\tau^a$. For most datasets, we can converge to the highest CA at $V_{\text{thr}}^a = 1.10$ and $\tau^a = 0.05$, where $ASR_o$ of +80\% for CIFAR-10, +75\% for GTSRB and +55\% for CIFAR-100 can be achieved. Interestingly, a higher attack effectiveness can be accomplished for CIFAR-100 ($ASR_p$ = +70\%) with simple power transformation. Besides, with the increase of attack hyperparameters, we observe consistent increase in both $ASR_p$ and $ASR_o$ yet degradations in CA. The CA degradation becomes more visible for CIFAR-10. For neuromorphic dataset N-MNIST, we observe that the Base CA is extremely close to Clean CA, while we can achieve 100\% ASR for all configurations. We conclude that by deliberately selecting attack hyperparameters, high CA and ASR can be achieved for all datasets and models.

\begin{table*}[h]
\caption{Performance analysis of \textit{BadSNN} across datasets and models.}
\label{tab:main_results}
\centering
\small
\begin{tabular}{llcccccc}
\toprule
Dataset & Model & Poison (\%) & Clean CA & $V_{\text{thr}}^t/\tau^t$ & Base CA & $V_{\text{thr}}^a/\tau^a$ & CA / $ASR_p$ / $ASR_o$ \\
\midrule

\multirow{3}{*}{CIFAR-10} 
& \multirow{3}{*}{ResNet-19} 
& \multirow{3}{*}{2} 
& \multirow{3}{*}{91.34} 
& \multirow{3}{*}{1.5 / 0.5} 
& \multirow{3}{*}{87.68} 
& 1.10 / 0.5 & 87.22 / 77.79 / 82.65 \\
& & & & & & 1.15 / 0.5 & 51.22 / 98.71 / 95.47 \\
& & & & & & 1.20 / 0.5 & 11.94 / 99.97 / 99.96 \\

\midrule

\multirow{3}{*}{GTSRB} 
& \multirow{3}{*}{VGG-16} 
& \multirow{3}{*}{5} 
& \multirow{3}{*}{96.05} 
& \multirow{3}{*}{1.5 / 0.8} 
& \multirow{3}{*}{93.02} 
& 1.10 / 0.5 & 92.57 / 39.81 / 75.59 \\
& & & & & & 1.15 / 0.5 & 91.43 / 54.81 / 79.75 \\
& & & & & & 1.20 / 0.5 & 87.43 / 71.92 / 85.08 \\

\midrule

\multirow{3}{*}{CIFAR-100} 
& \multirow{3}{*}{VGG-16} 
& \multirow{3}{*}{1} 
& \multirow{3}{*}{71.85} 
& \multirow{3}{*}{1.5 / 0.5} 
& \multirow{3}{*}{65.19} 
& 1.10 / 0.5 & 60.91 / 73.88 / 57.20 \\
& & & & & & 1.15 / 0.5 & 55.04 / 82.16 / 64.28 \\
& & & & & & 1.20 / 0.5 & 44.98 / 86.19 / 72.88 \\

\midrule

\multirow{3}{*}{N-MNIST} 
& \multirow{3}{*}{N-MNIST Net} 
& \multirow{3}{*}{3} 
& \multirow{3}{*}{96.06} 
& \multirow{3}{*}{1.5 / 0.5} 
& \multirow{3}{*}{95.10} 
& 1.10 / 0.5 & 94.06 / 100 / -- \\
& & & & & & 1.15 / 0.5 & 93.04 / 100 / -- \\
& & & & & & 1.20 / 0.5 & 92.19 / 100 / -- \\

\bottomrule
\end{tabular}
\end{table*}

\begin{table*}[h]
\caption{Baseline comparison and robustness analysis of \textit{BadSNN}. Highest ASR values are in bold.}
\label{tab:defense}
\centering
\footnotesize
\begin{tabular}{llcccccc}
\toprule
Dataset & Attack & Before & Fine-Tuning & CLP & ANP & TSBD & NAD \\
& & CA(\%) / ASR(\%) & CA(\%) / ASR(\%) & CA(\%) / ASR(\%) & CA(\%) / ASR(\%) & CA(\%) / ASR(\%) & CA(\%) / ASR(\%) \\
\midrule

\multirow{5}{*}{CIFAR-10}
& BadNet & \textbf{91.05 / 100} & 56.66 / 34.0 & 92.25 / 34.94 & 82.38 / 19.25 & 82.11 / 9.4 & 59.97 / 13.07 \\
& Blend  & 89.88 / 99.69 & 84.7 / 16.85 & \textbf{89.33 / 99.22} & 72.89 / 49.45 & 85.15 / 17.73 & 81.71 / 8.6 \\
& WaNet  & 88.24 / 99.58 & 77.76 / 22.48 & 86.96 / 99.01 & 68.43 / 25.89 & 87.11 / 17.31 & 84.83 / 15.7 \\
& \textbf{\textit{BadSNN}} & 87.22 / 82.66 & \textbf{82.25 / 84.55} & 83.83 / 72.59 & \textbf{28.93 / 98.43} & \textbf{71.27 / 82.61} & \textbf{87.02 / 81.89} \\

\midrule

\multirow{5}{*}{GTSRB}
& BadNet & 87.09 / 80.80 & 70.93 / 2.92 & 86.59 / 76.43 & 65.89 / 5.26 & 87.58 / 3.02 & 88.42 / 3.41 \\
& Blend  & 93.29 / 96.71 & \textbf{93.52 / 6.25} & 93.29 / 96.71 & 88.61 / 80.10 & 92.05 / 73.81 & 90.64 / 6.94 \\
& WaNet  & \textbf{95.99 / 99.78} & 16.53 / 2.68 & 95.99 / 99.83 & 2.14 / 0.0 & 93.49 / 3.00 & 31.94 / 2.90 \\
& \textbf{\textit{BadSNN}} & 87.43 / 84.99 & 38.70 / 1.69 & \textbf{3.33 / 100.0} & \textbf{11.05 / 99.80} & \textbf{87.65 / 84.99} & \textbf{70.10 / 9.86} \\

\midrule

\multirow{5}{*}{CIFAR-100}
& BadNet & \textbf{70.49 / 99.98} & 9.73 / 0.51 & \textbf{70.13 / 98.07} & 1.05 / 0.0 & 5.48 / 0.12 & 10.63 / 0.08 \\
& Blend  & 71.08 / 98.49 & 11.41 / 0.0 & 71.13 / 90.88 & 1.09 / 0.0 & 5.18 / 2.61 & 11.23 / 0.05 \\
& WaNet  & 70.24 / 95.12 & 11.24 / 0.89 & 70.24 / 95.19 & 0.94 / 4.83 & 4.56 / 0.0 & 9.96 / 0.78 \\
& \textbf{\textit{BadSNN}} & 61.18 / 74.23 & \textbf{6.27 / 16.49} & 66.73 / 6.66 & \textbf{1.07 / 100.0} & \textbf{2.45 / 53.09} & \textbf{5.9 / 7.4} \\

\bottomrule
\end{tabular}
\end{table*}

\begin{figure*}[t]
    \centering

    \begin{subfigure}{0.32\linewidth}
        \centering
        \includegraphics[width=\linewidth]{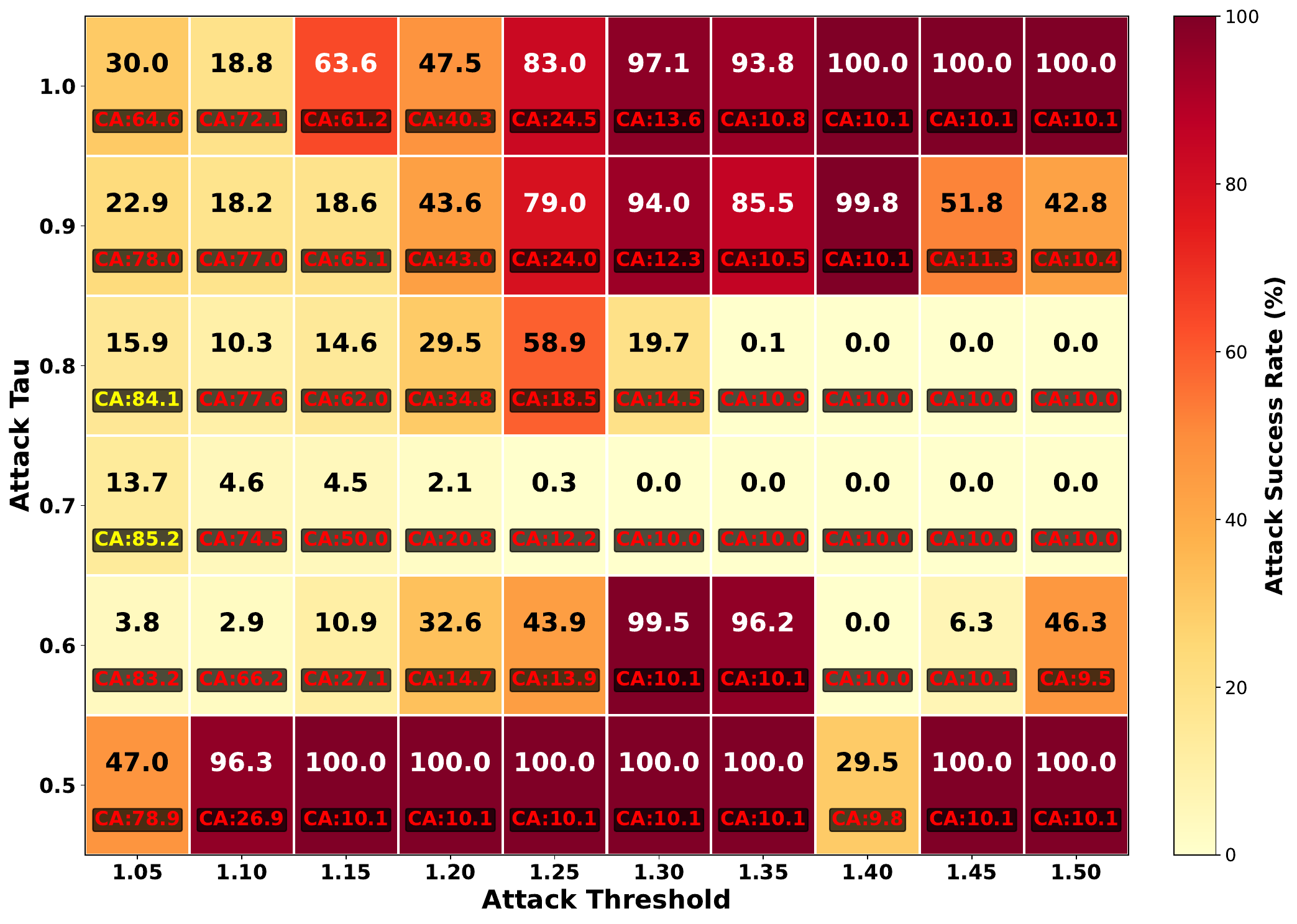}
        \caption{$V_{\text{thr}}^t = 0.8$, $\tau^t = 0.5$}
    \end{subfigure}
    \hfill
    \begin{subfigure}{0.32\linewidth}
        \centering
        \includegraphics[width=\linewidth]{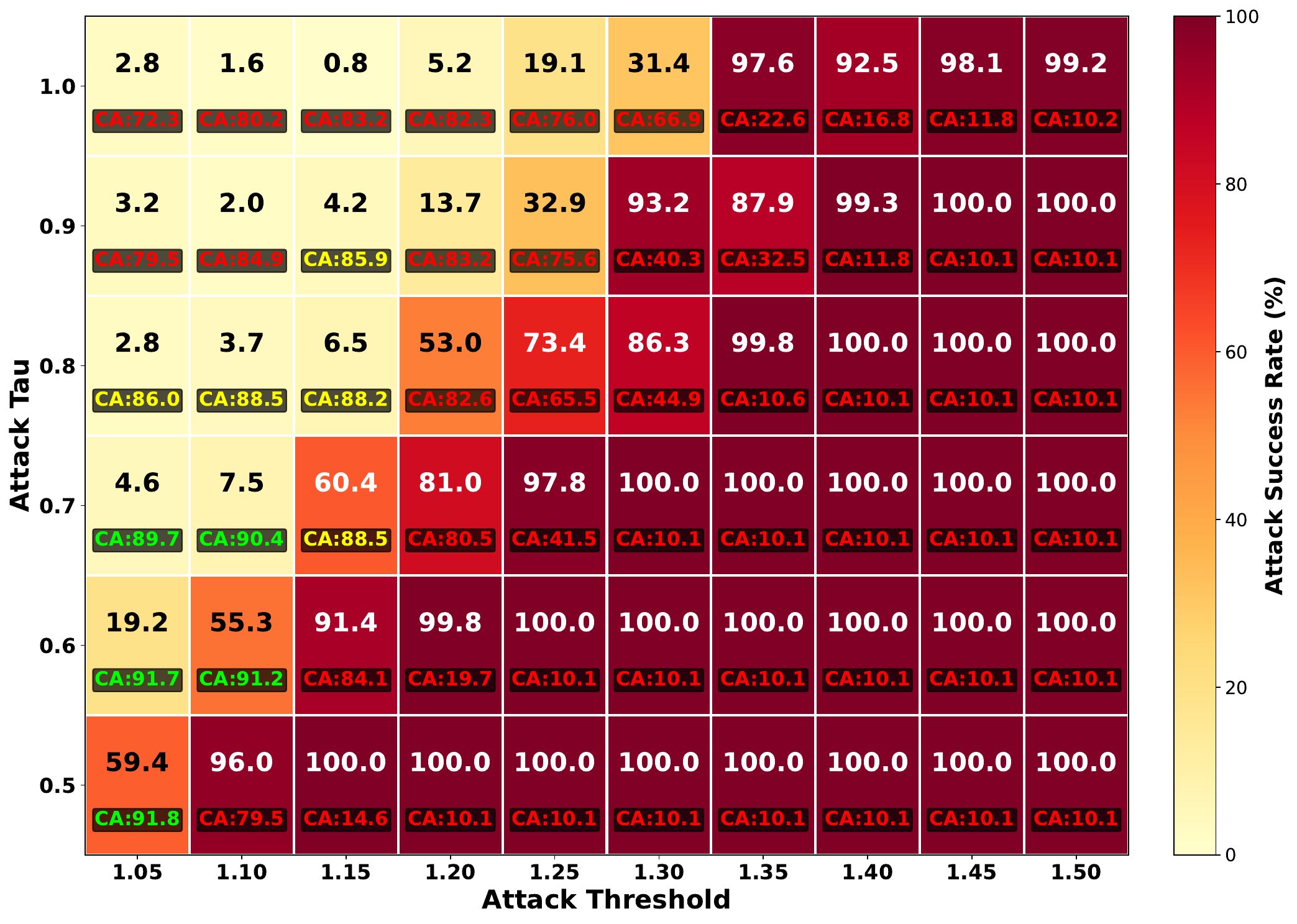}
        \caption{$V_{\text{thr}}^t = 1.25$, $\tau^t = 0.5$}
    \end{subfigure}
    \hfill
    \begin{subfigure}{0.32\linewidth}
        \centering
        \includegraphics[width=\linewidth]{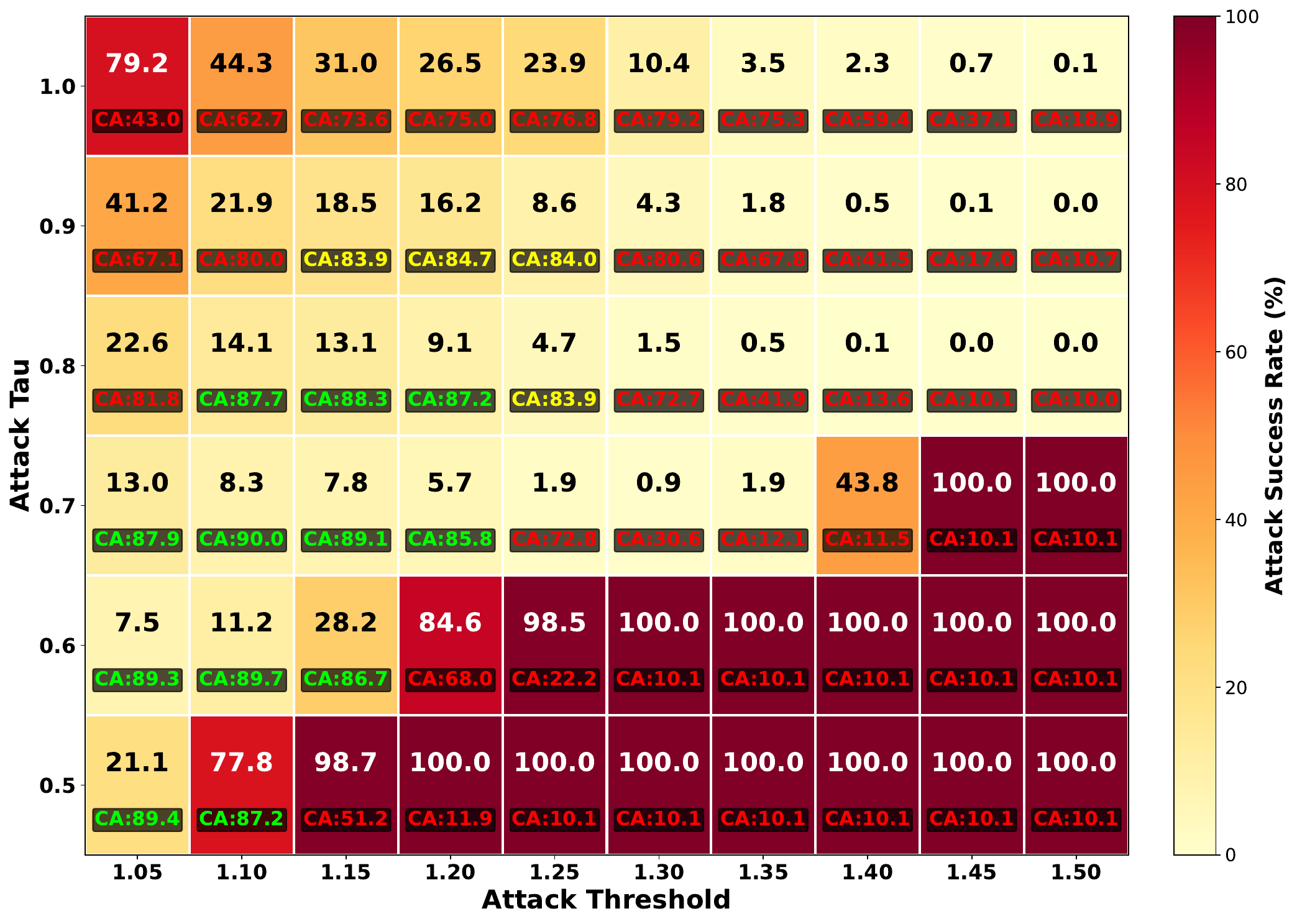}
        \caption{$V_{\text{thr}}^t = 1.5$, $\tau^t = 0.5$}
    \end{subfigure}

    \caption{CA/ASR heatmaps for different $V_{\text{thr}}^t$ and $\tau^t$.}
    \label{fig:heatmaps}
\end{figure*}

\begin{figure}[t]
    \centering
    \begin{subfigure}[b]{0.48\linewidth}
        \centering
        \includegraphics[width=\linewidth]{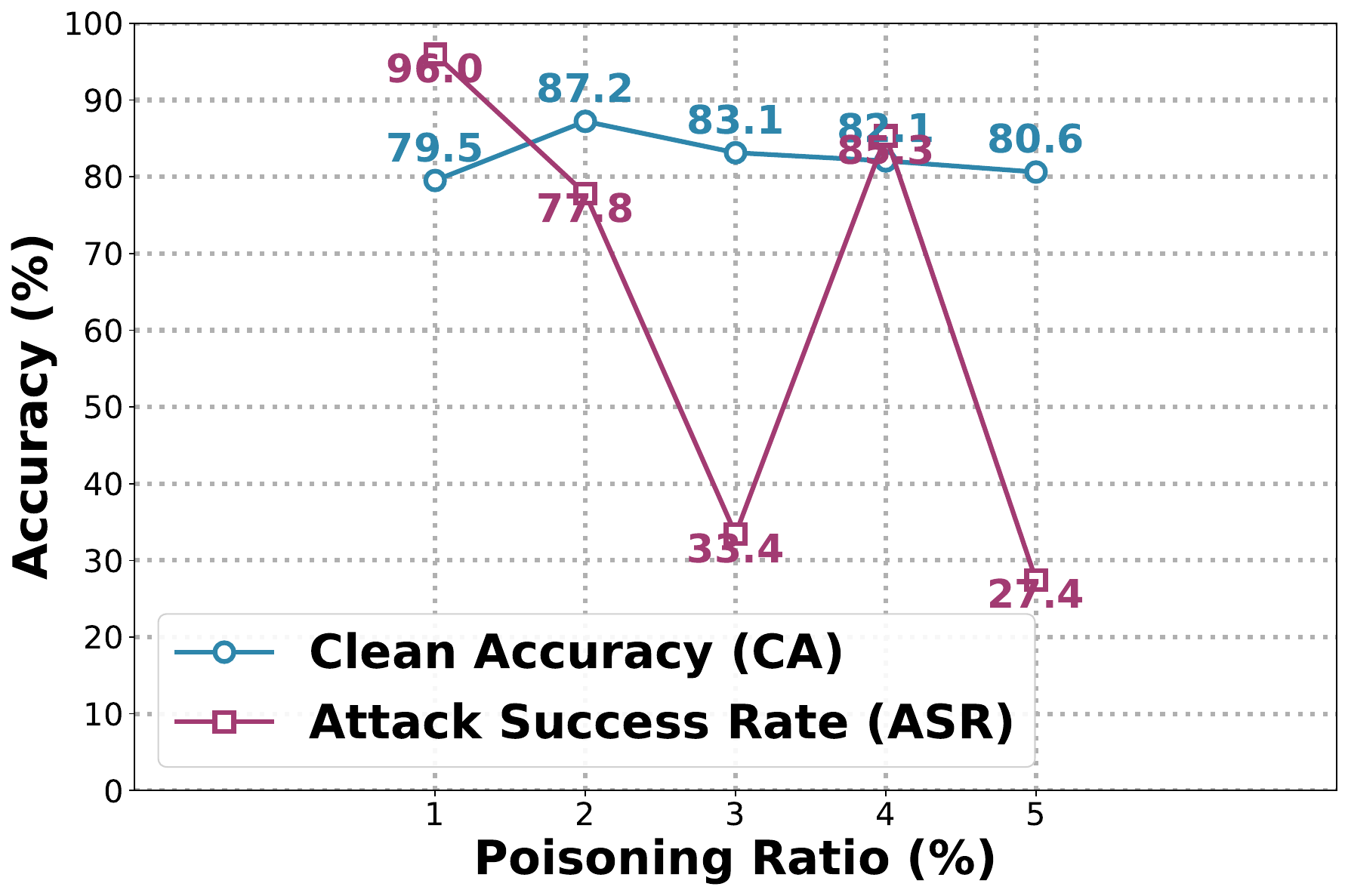}
        \caption{Performance for different poisoning ratios}
        \label{fig:poisoning_ratio}
    \end{subfigure}
    \hfill
    \begin{subfigure}[b]{0.48\linewidth}
        \centering
        \includegraphics[width=\linewidth]{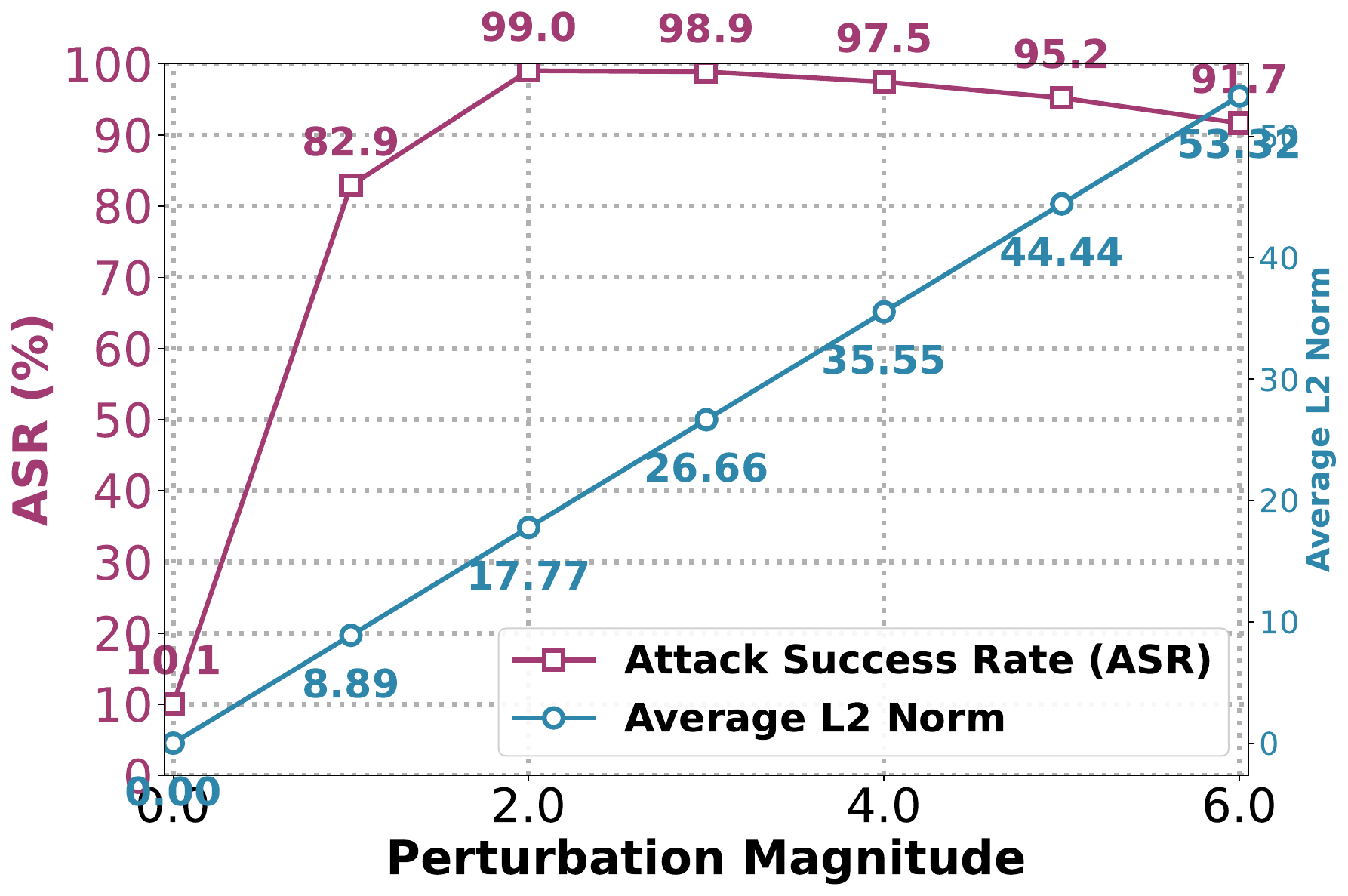}
        \caption{Performance for different perturbation magnitudes}
        \label{fig:magnitude}
    \end{subfigure}

    \caption{Attack effectiveness analysis for different poisoning ratios and perturbation magnitudes.}
    \label{fig:backdoor_parameters}
\end{figure}

\begin{figure} [t]
    \centering
    \includegraphics[width=0.7\linewidth]{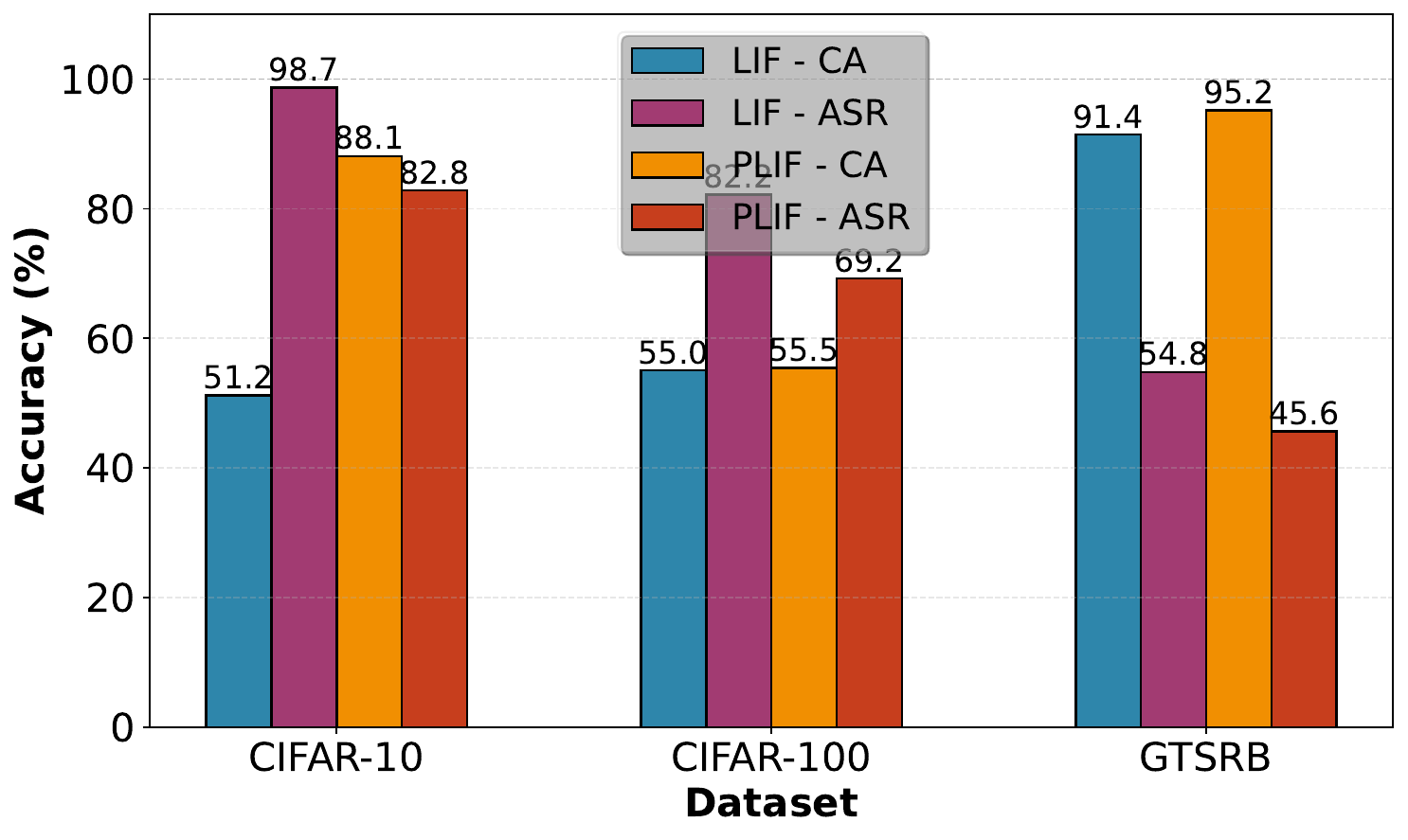}
    \caption{LIF vs PLIF}
    \label{fig:lif_vs_plis}
\end{figure}

\begin{figure} [t]
    \centering
    \includegraphics[width=0.7\linewidth]{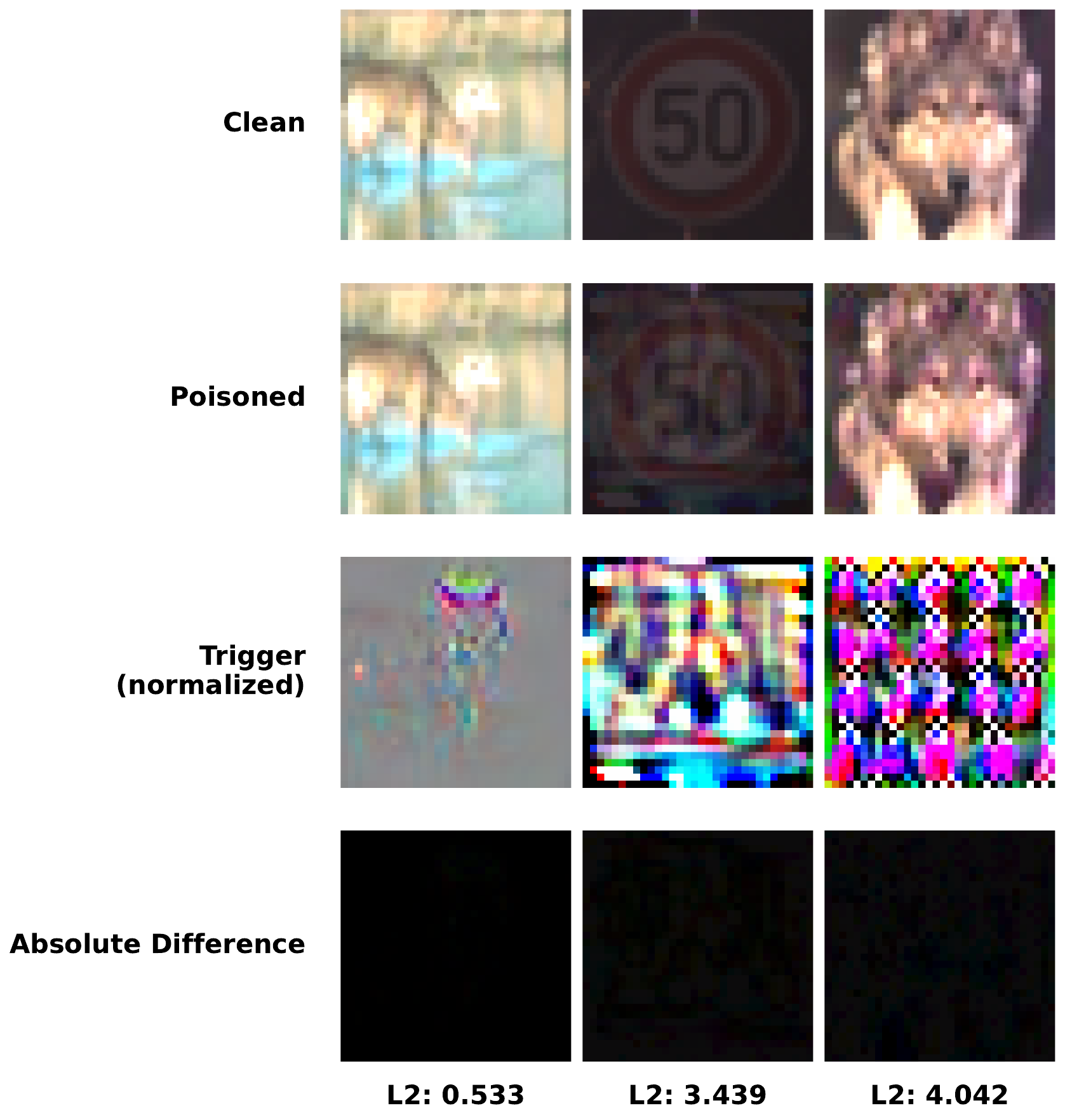}
    \caption{Different images from CIFAR-10, GTSRB, and CIFAR-100 with their corresponding triggered versions.}
    \label{fig:trigger}
\end{figure}

To demonstrate the effectiveness of \textit{BadSNN}, we summarize the results for different datasets and models when they are trained with given malicious hyperparameters ($V_{\text{thr}}^t$ and $\tau^t$) and evaluated under different attack hyperparameters ($V_{\text{thr}}^a$ and $\tau^a$) in Table  \ref{tab:main_results}. 

An acceptable accuracy degradation of Base CA from Clean CA can be observed across all datasets, with CIFAR-10 and GTSRB experiencing modest drops of approximately 3--4\%, while CIFAR-100 exhibits a larger gap of $6.7\%$. The CA and ASR vary considerably for different $V_{\text{thr}}^a$ and $\tau^a$, revealing a trade-off between clean utility and attack effectiveness. For most datasets, we identify the best trade-off at $V_{\text{thr}}^a = 1.10$ and $\tau^a = 0.5$, where the model retains high CA while achieving $ASR_o$ of $+$80\% for CIFAR-10, $+$75\% for GTSRB, and $+$55\% for CIFAR-100. As $V_{\text{thr}}^a$ increases beyond this point, ASR approaches near-perfect levels but CA degrades sharply. An interesting cross-dataset pattern emerges when comparing $ASR_p$ and $ASR_o$. For CIFAR-100, the simple power transformation yields higher ASR ($ASR_p = +$70\%) than the optimized trigger ($ASR_o = +$57\%) at $V_{\text{thr}}^a = 1.10$. In contrast, the optimized trigger consistently outperforms the power transformation for CIFAR-10. 

As of the neuromorphic dataset N-MNIST, we observe that the Base CA is extremely close to Clean CA, while 100\% $ASR_p$ is achieved across all attack configurations. It suggests that spike-native neuromorphic data is particularly susceptible to spike-level manipulation, as the temporal spike representation already operates in the same domain that \textit{BadSNN} exploits. We conclude that by deliberately selecting attack hyperparameters, high CA and ASR can be achieved across all datasets and models.

\noindent\textbf{Takeaway 1.} \textit{BadSNN} achieves a favorable trade-off between CA and ASR across diverse datasets and architectures, with moderate attack hyperparameters ($V_{\text{thr}}^a \approx 1.10$--$1.15$) offering the best balance, and neuromorphic data being inherently more vulnerable to the dual spike learning.

\subsection{Attack Robustness Analysis}

In this section, we compare our proposed \textit{BadSNN} with three baseline attacks and analyze its robustness against five state-of-the-art backdoor mitigation methods, summarized in the Table \ref{tab:defense}. While comparing the results of \textit{BadSNN} with baseline attacks and defenses, we report the best results found in the Table \ref{tab:main_results}. We also report $ASR_o$ for CIFAR-10 and GTSRB, and $ASR_p$ for CIFAR-100.

\noindent\textbf{Pruning-based defenses (CLP and ANP).}  CLP shows ineffectiveness against most attacks, including \textit{BadSNN}. ANP degrades the ASR to a meaningful extent for all baseline attacks but fails against \textit{BadSNN} across all three datasets. One interesting observation is that for \textit{BadSNN} the clean utility has been degraded a lot for ANP but the ASR remains strong for all of the datasets. Since \textit{BadSNN} embeds the backdoor through the dual-spike learning paradigm rather than through explicit trigger patterns, backdoor-related neurons are more thoroughly entangled with clean neurons. This entanglement prevents pruning-based methods from isolating and removing backdoor-specific neurons without simultaneously destroying the model's clean utility.

\noindent\textbf{Fine-tuning-based defenses (Fine-Tuning, TSBD, and NAD).} After performing vanilla fine-tuning on all attack models, we observe that the ASR is substantially reduced for all three baseline attacks across the three datasets. In contrast, our proposed \textit{BadSNN} remains effective on CIFAR-10 with $ASR_o = 84.55\%$ post fine-tuning. For GTSRB and CIFAR-100, the vanilla fine tuning causes substantial degradation in both the CA and ASR, hindering any meaningful mitigation. TSBD successfully mitigates all three baseline attacks, reducing their ASR below 20\% in most cases. However, \textit{BadSNN} remains robust against TSBD across all three datasets, retaining $ASR_o$ of 82.61\% on CIFAR-10 and 84.99\% on GTSRB. For CIFAR-100, while the ASR degrades slightly, the CA degradation is more noticeable. TSBD relies on activeness-aware fine-tuning to identify and suppress neurons with anomalous activation patterns. Because \textit{BadSNN}'s dual-spike learning causes backdoor neurons to exhibit similar activeness profiles to clean neurons across both nominal and malicious spike regimes, TSBD lacks a discriminative signal to selectively suppress them. NAD removes backdoor effects from all three baseline attacks across all datasets. Against \textit{BadSNN}, NAD is effective on GTSRB ($ASR_o = 9.86\%$) and CIFAR-100 ($ASR_o = 7.4\%$), but fails on CIFAR-10 where \textit{BadSNN} preserves $ASR_o = 81.89\%$. NAD relies on attention distillation between a clean teacher and the backdoored student to suppress trigger-specific attention patterns. Since \textit{BadSNN} does not produce localized trigger-specific feature maps, the distillation process cannot identify consistent attention discrepancies on CIFAR-10, where the attack's spike-level embedding is most deeply integrated with the learned representations.

\noindent\textbf{Takeaway 2.} \textit{BadSNN} exhibits strong resilience against both pruning-based and fine-tuning-based defenses because its dual-spike learning paradigm entangles backdoor behavior with clean representations, making it the only attack that consistently maintains high ASR across the majority of defense settings. This experiment showcases the benefit of spike poisoning based backdoor learning rather than traditional trigger poison-based backdoor learning.

\subsection{Ablation Studies}
\vspace{-3mm}
\subsubsection{Attack performance analysis for different $V_{\text{thr}}^t$ and $\tau^t$}

To gain deeper insight into the selection of malicious hyperparameters during backdoor training and the choice of attack hyperparameters during inference, we train spiking ResNet-19 on CIFAR-10 using different values of $V_{\text{thr}}^t$ and $\tau^t$, and analyze the CA and $ASR_p$ for varying $V_{\text{thr}}^a$ and $\tau^a$ in each scenario. The results are summarized as heatmaps in Figure \ref{fig:heatmaps}. For all experiments, we set $V_{\text{thr}}^n = 1.0$ and $\tau^n = 0.5$. When $V_{\text{thr}}^t < V_{\text{thr}}^n$, such as in the first scenario where $V_{\text{thr}}^t = 0.8$, it becomes very difficult to find a suitable pair $(V_{\text{thr}}^a, \tau^a)$ that yields both high CA and high ASR. Conversely, when $V_{\text{thr}}^t > V_{\text{thr}}^n$ with a small difference, it is possible to find such a pair. Yet, because the malicious spikes are leaning towards the nominal spikes, achieving strong attack performance results in accepting a slightly lower CA. When $V_{\text{thr}}^t = 1.5$, the regions of high CA and high ASR become more visibly separated, making it easier to find a middle ground where both CA and ASR are satisfactory. Based on such analysis, we infer that malicious hyperparameters should be chosen such that $V_{\text{thr}}^t > V_{\text{thr}}^n$ with a sufficiently large difference between them to ensure effective attack behavior without excessively degrading clean accuracy.

\noindent\textbf{Takeaway 3.} The malicious backdoor training hyperparameters must be set sufficiently above the nominal hyperparameters to create a well-separated spike distribution; $V_{\text{thr}}^t = 1.5$ with $V_{\text{thr}}^n = 1.0$ provides the most favorable and flexible CA \& ASR trade-off landscape.

\subsubsection{Attack performance for different poisoning ratios} We analyze the performance of \textit{BadSNN} under different poisoning ratios on CIFAR-10 using the spiking ResNet-19 model and observe the corresponding variations in CA and $ASR_p$ shown in Figure \ref{fig:poisoning_ratio}. Poison ratios of 1\%, 3\% and 5\% present better attack performance with both high CA and ASR, while poison ratios of 2\% and 4\% experience significant ASR drops despite preserved clean accuracy. The above results suggest that the relationship between poisoning ratio, CA and $ASR_p$ is highly non-linear orthogonal to experimental observation in state-of-the-art attacks \cite{gu2019badnets,chen2017targeted,nguyen2021wanet,turner2018clean}.

\noindent\textbf{Takeaway 4.} The relationship between poisoning ratio and attack effectiveness is non-linear in \textit{BadSNN}, with odd ratios (1\%, 3\%, 5\%) yielding consistently better performance, highlighting the need for empirical calibration rather than naive ratio maximization.

\subsubsection{Perturbation magnitude analysis} We analyze the $ASR_o$ of CIFAR-10 on the spiking ResNet-19 model for various perturbation magnitudes generated by $\mathcal{T}_o$ presented in Figure \ref{fig:magnitude}. We also report the L2 norm between the clean and triggered images with perturbation. We observe a sharp increase in $ASR_o$, reaching 99\% with a perturbation magnitude of only 2.0. However, as the perturbation magnitude continues to increase, the L2 norm also grows approximately linearly leading to rising sample distortion. The clean image, the triggered image obtained by adding perturbations from $\mathcal{T}_o$, and the normalized trigger perturbation are illustrated in Figure \ref{fig:trigger}.

\noindent\textbf{Takeaway 5.} The optimized trigger achieves near-perfect ASR at moderate perturbation magnitudes ($2.0$), beyond which distortion increases with larger perturbation that are largely unnecessary and counterproductive.

\subsubsection{LIF vs. PLIF} All the previous experiments are based on the assumption that spiking neurons adopt the LIF model. However, a recent study named Parametric LIF (PLIF) \cite{fang2021incorporating} adopts a learnable $\tau$ for spiking networks. 
To evaluate the effectiveness of \textit{BadSNN} on PLIF, we vary only $V_{\text{thr}}$ for nominal and backdoor training, 
setting $V_{\text{thr}}^t = 1.5$ and $V_{\text{thr}}^a = 1.15$ for both LIF and 
PLIF models. The performance differences between LIF and PLIF are shown in 
Figure \ref{fig:lif_vs_plis}. At $V_{\text{thr}}^a = 1.15$, LIF achieves lower 
CA for CIFAR-10, while PLIF maintains considerably higher CA, with both models 
demonstrating strong $ASR_p$. Both CA and ASR, however, are comparable between LIF and PLIF models for CIFAR-100 and GTSRB. 

\noindent\textbf{Takeaway 6.} PLIF's learnable $\tau$ improves clean accuracy retention under attack conditions but does not mitigate the backdoor itself, confirming that \textit{BadSNN} is effective against both fixed and learnable spiking neuron models.

%\vspace{-5mm}

%\input{sec/5_discussion}

\section{Conclusion} \label{sec:conclusion}

\vspace{-1mm}

In this paper, we propose \textit{BadSNN}, a novel backdoor attack on spiking 
neural networks that exploits hyperparameter variations of spiking neurons to 
embed backdoor behavior. We further propose a trigger optimization process to 
enhance attack performance while maintaining imperceptibility. \textit{BadSNN} 
offers two key advantages over conventional data poisoning-based attacks: (i) 
it eliminates the need for input data manipulation, providing greater 
stealthiness, and (ii) it demonstrates superior robustness against 
state-of-the-art backdoor mitigation techniques. This work motivates the 
development of more effective defenses for spiking neural networks.

%\newpage

\bibliographystyle{unsrt}
\bibliography{acmart}

\end{document}